\documentclass[final,abstracton]{article}
\usepackage{arxiv}

\usepackage[english]{babel}

\usepackage{graphicx}
\usepackage{wrapfig}

\usepackage{etoolbox}%
\usepackage{microtype}

\usepackage[pdftex,unicode=true,breaklinks=true,colorlinks=false,linkcolor=black,pdfborder={0 0 0},pdfdisplaydoctitle=true,pdftitle={Data science on industrial data – Today’s challenges in brown field applications},pdfauthor={Tilman Klaeger}]{hyperref}
\hypersetup{final}

\usepackage[style=numeric, uniquelist=false,uniquename=false,giveninits=true, url=false,isbn=false,backend=biber,maxbibnames=25,sorting=none,bibencoding=inputenc, maxcitenames=2, autocite=inline, date=year, sortlocale=auto, natbib=true, ibidtracker=false]{biblatex}

\usepackage[output-decimal-marker=comma,retain-unity-mantissa=false,separate-uncertainty=true]{siunitx}
\title{Data science on industrial data – Today’s challenges in brown field applications}
\author{Tilman Klaeger, Sebastian Gottschall, Lukas Oehm\\
	Fraunhofer Institute for Process Engineering and Packaging (IVV)\\
	Division Machinery and Processes\\
	Dresden, Germany \\
	{tilman.klaeger@ivv-dd.fraunhofer.de}
}

\begin{document}

\maketitle

\begin{abstract}
Much research is done on data analytics and machine learning. In industrial processes large amounts of data are available and many researchers are trying to work with this data. In practical approaches one finds many pitfalls restraining the application of modern technologies especially in brown field applications. With this paper we want to show state of the art and what to expect when working with stock machines in the field.

A major focus in this paper is on data collection which, in real scenarios, can be more cumbersome than most people might expect. 
But also data quality for machine learning applications is a major challenge once leaving the laboratory. 
In this area one has to expect the lack of systematic semantic description of the data as well as very little ground truth being available for training and verification of machine learning models.
A last challenge when working with live industrial data is IT security and securely passing data through various types of firewalls.

\end{abstract}

\section{Introduction}

Many studies show the possibilities for cyber-physical systems in contexts of Industry 4.0. A lot of projects presented are using the latest technologies or circumvent too direct interaction with existing infrastructure by applying Industrial Internet of Things (IIoT) architectures.

Very little is written about working with machines in brown field applications and doing data science on data extracted from those machines.
This paper represents a personal perspective on data science on data collected in the field. Some findings generalize very well and are well known when talking to other people in the area of research but only few are actually written down. Aim of this paper is therefore to help people new to the subject to understand the current situation. Looking at typical machine life-cycles of 10 to sometimes more than 20 years this situation will change but slowly. Adding the limitations current automation platforms provide also machines developed today will not always contain all features proclaimed in the scientific community.

There are major differences between industrial and academic data science \cite{Menzies2011inductivesoftwareengineering}. Some topics noted are currently under intensive research and development whereas other topics seem to be unresolved at the moment. Various research opportunities are well known \cite{Bordeleau2018BusinessIntelligenceIndustry} and many challenges are still open for brown field as well as for green field applications \cite{Wang2018IndustrialBigData}.
In this paper we will mostly focus on brown field applications and summarize existing literature as well as personal experience gained in the last years. 

\section{Raw Data Collection on the field}

A data science project in an industrial context usually starts at the field level to collect data from existing machines and sensors adding some kind of edge or IIoT device.
Generally spoken there are some different approaches from OPC UA, field bus integration to debugging interfaces and external IoT-Devices to read the data.
A combination of those techniques is usually possible but causing more efforts in data acquisition and merging.

\subsection{OPC UA}
Many machine controllers (programmable logic controller, PLC) set up in the last years support the standard OPC UA as machine-to-machine interface.
It not only provides a standardized protocol but also adds semantics to the data in order to have a description of the data available.
Depending on the configuration of the OPC UA server in the controller reading and writing selected or all values of the PLC is possible.
For reading values not only polling – periodic reading of the values – is possible but also a subscription mode.
In this mode new data will be pushed to the subscribing data acquisition client at the moment the values change.
Connecting those to an edge device or even to a complete machine network should therefore be easy. 
With OPC UA TSN data transfer over Ethernet will even be possible in hard real time and thus providing a good solution  for time critical applications \cite{Gogolev2019TSNTrafficShaping}.
The key is the application of »Time Sensitive Networks« (TSN) to warrant configurable time slots just for the transfer of real-time data like OPC UA data.

One major drawback for current applications is the time needed to read the data which has been shown by \cite{Wazny2015Configurationperformancetest} and correlates well with our own experience.
Reading a variable from a Soft-PLC in polling-mode can take up to \SI{20}{ms}.
Sometimes over 1000 parameters need to be read from the machine. 
This will severely slow down the process of data acquisition.
If this turns out being too slow it is possible to build a cache in the controller and read the data as an array. But this will result in the loss of one of OPC UA’s main features: Semantic description of the data. 
With good implementations more performance is possible \cite{Profanter2019OPCUAROS} but you cannot always expect to have those in the machine chosen for the project. 

Another major pitfall in OPC UA is the incomplete implementation of the standard in many PLCs \cite{Weyer2015IndustryStandardizationcrucial}. At CERN this caused engineers to implement automatic testing of OPC UA implementations as there is no guarantee for a complete and bug free implementation of the standard \cite{Farnham2011AUTOMATEDTESTINGOPC}.
Polling data seems to be available always, but you cannot expect subscription mode to work.
If this is available, one may experience sudden outages with no more data arriving at the data acquisition tool.

This may slowly be changing with newer PLCs being available to machine manufacturers and, after thorough testing, in the field replacing older machines.
But even for latest machines in the machine manufacturer’s laboratory special solutions are built to access the data as current standards are not suitable for all needs \cite{Kammerer2019AnomalyDetectionsManufacturing}.

\subsection{Integration in field bus}
If data in higher frequency is needed than OPC UA can provide or OPC UA is not available one may try to read the field bus directly.
Here one will face different protocols by different vendors \cite{Mahalik2009Extendingfieldbusstandards, Bader2018EnablingCyberPhysicalSystems}.
When looking at a manufacturing line, the machines may even use different protocols making the process of data acquisition even more difficult.
The proposed solution in many edge devices providing field bus support is to add those devises as an additional slave in the system and have the PLC write the wanted data to this new slave.
It therefore needs to provide the same protocol as the existing field bus.
For widely used protocols there are solutions available but looking at older systems like the fiber-optic based Sercos 2 or 1970s Arcnet solutions are hard to find if available at all.

Depending on the complexity and age of a machine often there is no possibility to add a new edge device as a slave in the automation bus. Automation environments for programming the PLC changed over the years and the old version may not be compatible with current operating systems.
Also reprogramming with the control software’s source-code available at the machine manufacturer may overwrite minor but crucial improvements or fixes done in the field.
This is completely independent of adding any additional, untested logic.
It seems not to be uncommon for a service technician to alter the machine code slightly to make up for some local problem and not always return the changes to the company’s server \cite{Khudyakov2018Versioncontrolsystem}.
Another source of diverging code bases is working with an internal code platform and adapting the resulting code to each customer.
Adding, even minor, features may make changes to old machines costly when an update to a new code base is needed \cite{Lettner2014CaseStudySoftware}.
Source code versioning  systems common in the software development are only slowly evolving in current PLC programming environments.
Especially for graphic programming from the IEC 61131 PLC languages version control with concurrent versions and merge strategies is a complex task \cite{Khudyakov2018Versioncontrolsystem}.
Thus, the usage of those techniques seems to be even more limited in a practical approach.
So accessing the data one often has to cope with »No changes to a running system« issued by the PLC programmer for a fair reason. 

However, one option to access the field bus is to build »sniffers« that capture traffic in a passive mode on the bus and decode the signals in an edge device.
Solutions like this can be purchased for some systems (to our knowledge for ProfiNet and EtherCAT), others worked on Profibus \cite{Mamo2017LegacyIndustryProfibus}.
To capture Modbus RTU, a 1970s protocol still widely used in the field, we developed a sniffer ourselves.
Of-the-shelf solutions may be difficult to build as there are always surprising solutions in the practice like usage of a bidirectional point-to-point RS422 connection instead of the more common asynchronous RS485 for Modbus data transfer.
Building a suitable gateway it is possible to create a Modbus to OPC UA mappings \cite{Tunkkari2018MappingModbusOPC} to address some brown field issues especially in the area of semantics.

Nevertheless, the accessible data is implementation dependent.
Notifications about machine faults e.\,g. may be sent over the field bus from PLC to the panel showing the human machine interface (HMI).
In some integrated systems however this may not be the case if panel and controller are one device. 
In other settings the HMI may be connected to the PLC with some other interface than the field bus.
Before implementation a thorough analysis of the available automation solution and it’s  interfaces is needed. 

\subsection{Using the debugging interface of the PLC}

Most PLCs have a so-called »Online Mode« during programming. 
What is available for development may also be used to read out data in production if the protocol is known.

With Siemens PLCs this is sometimes referred to as S7 protocol with a popular open source implementation called »Snap7«. Reading out the memory (process image) of a PLC is not difficult using this method \cite{Zheng2017DevelopmentRemotelyMonitoring}. 
But accessing the memory more or less raw requires to have a mapping available.
This in turn requires cooperation with the programmer of the machine controller.
Even using Siemens’ own »MindConnect Nano« designed for brown field data acquisition needs to know the addresses of the variables \cite{SiemensAGDivisionDigitalFactory2017MindSphereMindConnectNano}.
Depending on the project this may lead to conflicts of interest as some manufacturers are not able or not willing to share any details about their control program with their customers.
No regularities to whom the data belongs that is created in the machine are available at the moment.
For old machines the controller code or tools to view the code may not be available and thus getting access to the required mapping is impossible even if the machine manufacturer is willing to support the project.

If the addresses of the variables are not specifically set during programming those may easily change on software updates or largely differ between various machine revisions.
So this will not result in a plug and play solution for effortless adoption to different machines.

Similar to Siemens S7 protocol is Beckhoff PLC’s support for »ADS«. Reading out the names of variables is possible, provided you can set up an »ADS Route« which requires administrator permissions in the PLC.
Gaining this permission again depends on a trustful collaboration with the manufacturer or programmer of the machine.
The reading speed is not fast enough for high frequency data as each reading takes \SIrange{1}{4}{ms} response time and additional, unpredictable time for the Ethernet data transfer \cite{BeckhoffGmbHCoKG2003FieldbusNetworksWorkshop}.

Smaller PLC vendors may use different protocols.
Those are rarely documented, so available features are unknown and predicting the transfer speed is impossible.
Hence, reading the debugging interfaces on those PLCs is mostly not an option in one-time projects as it requires reverse engineering the protocol. 
Implementation seems only worth it when adaption of the system is planed for many machines.

\subsection{Third party device as sensor interface}
Another option to read data in the field is to set up a dedicated IIoT structure with devices connected directly to existing sensors or adding new ones \cite{Strauss2018EnablingPredictiveMaintenance, Burggraf2019Sensorretrofitcoffee}.
This provides major benefits as one has complete control over data formats and protocols being used to transfer the data. But on the other hand much data can not be acquired at all: There is no possibility to read internal states of the PLC or errors detected by the PLC. 
Also, reading out error codes, motor currents or lag errors are difficult if not impossible.

Some sensors like PT100 temperature sensors connected in a three- or four-wire circuit or sensors like resistance strain gauge connected with a bridge circuit can not be read out by two analog-digital converters at the same time without major efforts with custom designed circuits.

\subsection{Data collection from business intelligence}

Another source one might want to access is the area of business intelligence with data from manufacturing execution systems (MES) or enterprise resource management tools (ERP) like SAP. A lot of these tools may be called open, but what is deployed in the field is highly heterogeneous following different standards \cite{Wang2018IndustrialBigData}. 
Common standards are REST-APIs providing the data in Java Script Object Notation (JSON) over simple HTTP-Protocol. 
This is easy to read out, but of course requires adjusting the own data acquisition tools to work with the API. 
In similar manner data can be transferred over TCP/IP servers in XML-format. 

So implementing tools to read out and analyze this data is technically not difficult but requires lots of manual work. 
Sometimes also additional licenses have to be bought in order to open specific external interfaces.

\subsection{Summarizing data collection}

Each of these solutions has its benefits and drawbacks need to be carefully considered before setting up a system for data acquisition in brown field applications.
In most cases some data will not be readable with reasonable effort and costs.
Efforts are put towards standardization but for current projects one has to deal with tedious work not bringing scientific progress in order to acquire data for research.

\section{Data quality}

\subsection{Lack of ground truth to train models}

Producing lots of (raw) data is easy in industrial processes \cite{Oliveira2019IndustryFocusedData}. What is much more difficult is obtaining a reasonable ground truth to build machine learning models, which is addressed less frequently \cite{Obdenbusch2018Referenzarchitekturfuercloudbasiertes}. Especially classification data for fast running processes is difficult to produce \cite{Klaeger2019Usinganomalydetection}. Different methods used are very suitable for research but, to our opinion, are not fully adoptable to live industrial processes. 

One approach is to provoke longer lasting faults and classifying the resulting data \cite{Zurita2015Diagnosismethodbased, Klaeger2017LernfahigeBedienerassistenzfur, Brecher2017Optimizedstateestimation}. Not looking for classification but for anomalies much data is needed, which still can be produced in a local experiment, as shown by \cite{Kammerer2019AnomalyDetectionsManufacturing}.
Building models to detect longer lasting anomalies like wear or malfunctions in bearings is much different.
Looking at steadier processes than discrete manufacturing some problems are easier to solve, even though more machines need to be available for a proper modeling \cite{Smart2013Comparingonetwo, Rapur2017ExperimentalTimeDomainVibrationBased}.
Detection anomalies in discontinuous processes as against steady ones is not much easier than classification as ground truth often needs to be determined for single products to validate the generated models.

When working with continuous or batch processes like food or chemical production one may be able to work on a lot of historical data.
Data-Logging in this area is much more common already since data needs to be available in a  central control room \cite{Mersch2011GemeinsamkeitenundUnterschiede, Muller2018Processindustriesdiscrete}.
Data in such processes comes at a slower rate than those originating in discrete processes.
If data is used to detect rare and hard to find events the ground truth may also not be available.
One such event is the so-called fouling which will produce stains especially in heat exchangers of industrial processes. 
If this occurs operators notice it has happened with some time delay not knowing when it actually got severe.
Detecting such an event is hard and thus application of new methods like machine learning seems suitable.
Training those models is hard as no secured ground truth is available to accompany the historic data.

One way to build a ground truth may be to manually watch the process. 
This has some severe drawbacks: 
First it is very labor intensive to monitor the process, especially if events do not occur very frequently.
On other processes the processing speed is simply to high for a human to monitor manually \cite{Klaeger2019Usinganomalydetection}.

Building models to predict events in the (near) future has one severe benefit as opposed to classification or anomaly detection in historic data:
Annotation of training data will happen automatically based on events happening, at least as long as the events can be detected and recorded.
This technique can of course be used for weather forecasts but has also been proven to work in industrial settings like production of plastic films \cite{Kohlert2015Multisensorydataanalysis}.

When setting up projects the focus at an early state should therefore look towards the ground truth data acquisition.

\subsection{Obtainable training data quality}

A major concern for all models and especially machine learning models is generalization:
»Will the model work on new data as well as on collected training data?«
One key issue are spurious relationships leading to over-fitted models \cite{LHeureux2017MachineLearningBig}.
During training those models »memorize« all correlations independent of causality and thus perform well on the provided data but not future data.
Statistically it can easily be proved, that storks deliver babies \cite{Matthews2000StorksDeliverBabies}.
There is no causality behind this finding, but the statistical model does show this clearly.
For offline-learning applying a good design of experiments is possible.
This allows to create statistical independent data for training of machine learning models.
Collecting live industrial data controlling the process is not always possible and thus the knowledge over generated data is smaller and may contain various spurious correlations hard to find and eliminate.

Closely related to this issue are concept drifts.
In this case the surrounding effects, not being part of the model, change over time.
Causes for this may be manual adjustments to the machine or wear in measuring equipment or electrical drives.
In the data collected for \cite{Klaeger2019Usinganomalydetection} we could notice such behavior:
The model working well on data for one month did not generalize for data of another month. 
Using different features and adjusting the model provided a better model. 
But it still can not be guaranteed to work in the future.

Determining the difference between a model drift or over-fitting on spurious correlations is one of the most difficult tasks in data science.
This turns out even more if one is not in complete control over the process being monitored and modeled.
Whereas there are many ideas on handling concept drift \cite{Lu2019LearningConceptDrift,Webb2017UnderstandingConceptDrift} finding spurious correlations can only be done with process and data understanding.

Other challenges are not only common in industrial processes but are to all measured data: 
Noise in the data and measurement errors like sensors with wrong or no reading at all.
Further one often has to cope with imbalanced data looking at many well produced parts and only very few with defects.

\section{Semantic description of available data}

Much research effort is put into semantic description of data to make it machine-readable and especially \mbox{-}understandable.
Efforts are going in  different directions using linked data approaches \cite{Graube2011LinkedDataIntegrating, Folmer2017BigundSmart}, description of devices like the Industry 4.0 reference architecture (RAMI 4.0) and similar device description approaches \cite{SmartmanufacturingReference, Goessling2014DeviceInformationModeling} and various other ways to describe sensors and connected machines \cite{Bunte2016Integratingsemanticsdiagnosis, Nilsson2018SemanticInteroperabilityIndustry, Dibowski2018SemanticDeviceSystem}.
Most promising seems to use semantics already designed in OPC UA.
It offers a variety of possibilities to annotate the data for special purposes like in application-to-application interfaces \cite{Graube2017InformationmodelsOPC}.
Even more valuable for future applications are so-called »Companion Standards«. 
Their aim is to provide a standardized data model for many applications. 
Models like this are not new like the »Weihenstephaner Standards« to provide for production data acquisition (PDA) in food processing plants and is implemented in many machines \cite{Flad2017AutomatischeGenerierungFertigungsManagementsystemen}.
For many applications companion standards are work in progress \cite{Graube2017InformationmodelsOPC}.
Many other data formats available including Modbus are transferred to OPC UA data models \cite{Seilonen2019OPCUAInformation, Tunkkari2018MappingModbusOPC}.
All these works can help to reduce the tedious manual work needed e.g. for feature selection in machine learning applications \cite{Ringsquandl2015SemanticGuidedFeatureSelection, Diedrich2017SemantikdurchMerkmale}. 

Going one step further semantic description and process modeling could be used for an easier if not automated feature engineering.
Some process modeling techniques are even sophisticated to generate PLC code automatically  \cite{VereinDeutscherIngenieure2015FormalisierteProzessbeschreibungInformationsmodell, Fay2017SemantischeInhaltefur, Arroyo2014DerivationDiagnosticModels}.
All these ideas are not new, but especially discrete manufacturing is very different in terms of process models opposed to continuous processes like chemical production  \cite{Mersch2011GemeinsamkeitenundUnterschiede}. 

In practical applications these ideas are not in use yet.
PLC programming is mostly manual work and the PLC programmer is supplied with various information sources, mainly informal ones \cite{Colla2009Designimplementationindustrial, Holowenko2017AssistenzSteuerungsentwicklungproduktionstechnischer}.
Process models are mostly not used when designing control programs.
The topic of incomplete OPC UA implementations arises again when looking to add semantic description to the data model in the automation framework. 
Tools provided today only have limited possibilities to create OPC UA data models, more sophisticated models are mostly not available.
In consequence one will face a number of more or less systematically named variables that need to be mapped to the aimed solution.

\section{IT infrastructure}

Heavily depending on IT infrastructures IT security gains rising attention even in the field level. A common technology is network segmentation using virtual local area networks (VLAN).
With this it is possible to assign each network jack or device a separate network independent of the actual network switch connected to. 
Those separate networks can be inter-connected by simple routers or more sophisticated firewalls depending on the specific need.
Some companies even go one step further having completely split networks for administrative tasks, machines and guests at their facility with separate internet connections for each network.

IT security at an infrastructure level is definitely needed as PLCs can have security flaws \cite{Karnouskos2011Stuxnetwormimpact, Klick2015InternetfacingPLCsnetwork}.
Software updates may mitigate the risk of security incidents but increase the risk of control software malfunctions. 
Especially in safety relevant areas proven to work or even certified solutions are not easy to update.
When using SoftPLCs, as many vendors offer, the PLC is running on a host operating system in a special real-time setup. 
Having the control system based on widely available operating systems brings features like easy integration into corporate networks and possibilities to install custom software for instance to log process data. 
Due to the update risks mentioned earlier the risk of security incidents is high.
Also, more integrated control hardware is often closely enmeshed with embedded operating systems like Microsoft Windows CE \cite{Colla2009SurveyMethodsTechnologies}.
IT security has to be in focus but nevertheless this can severely slow down data science projects when deploying local edge devices or IIoT hardware and connecting it to the internet.
For own research projects attached to existing infrastructure the security of own devices has to be kept in mind.

Some research is done in the area of IT security with the aim to provide guidelines for infrastructures suitable for modern cyber-physical systems \cite{Diemer2017SichereIndustrie40Plattformen, Fallenbeck2017ITSicherheitundCloud}.
In practice on will find different setups at every location to be coped with.

One major and lasting trend in IT is outsourcing. 
Increasing complexity of IT systems and moving resources to the cloud are causing many companies to have centralized or outsourced IT departments.
This may conflict with the need for flexible solutions on location.
Data acquisition is often needed in different areas: field level for sensors data up to the ERP to collect data about the currently manufactured product \cite{Perez2015CPPSArchitectureapproacha}. Sending data to external cloud storage may be even more cumbersome and, to our experience, may produce serious slow downs in the project.

At the end this may result in special solutions, we experienced ourselves not only once: Consumer-style networks using consumer-style routers for internet access. To have some freedom to operate those sometimes are even unknown to the central IT department.

\section{Conclusion}

Doing data science on data collected in industrial processes requires much tedious manual work at the moment.
It’s not only the often proclaimed »80\% of the time is needed for data preparation« but it easily extends by the time needed to just get the data to prepare.
There are interesting technologies to be integrated in future control software solutions.
But looking at typical lifetimes of machines these technologies will only emerge slowly. 

The difference in life cycles for software products and machines cast in iron and steel will be a major challenge even in the future.
So-called retrofits may be a solution and are provided by machine producers already to support their service business. Looking a scientific projects wanting to push to possible one step further waiting for the next retrofit is not always possible. Not wanting to wait to apply machine learning and other data scientific approaches to discrete manufacturing processes one has to cope with the challenges. So whenever setting up projects aiming at real data we recommend planning enough time for the simple but time-consuming task of data acquisition.

\printbibliography

\vfill

\begingroup
\setlength{\intextsep}{0pt}
\setlength{\columnsep}{3pt}
\parindent0pt

\begin{minipage}[l][1.2in]{\textwidth}
	\begin{wrapfigure}{l}{1.1in}
		\includegraphics[width=0.88in,height=1.1in,clip,keepaspectratio]{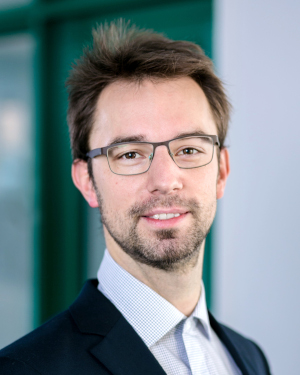}
	\end{wrapfigure}\par
	\textbf{Dipl.-Ing. Tilman Klaeger} studied mechatronics at the Technische Universit{\"a}t Dresden and started working at the Fraunhofer IVV in 2016 and is now in lead of the team data science. His major topic in research is machine learning on industrial data collected from packaging machines and processes.\par
\end{minipage}

\begin{minipage}[t][1.2in]{\textwidth}
	\begin{wrapfigure}{l}{1.1in}
		\includegraphics[width=0.88in,height=1.1in,clip,keepaspectratio]{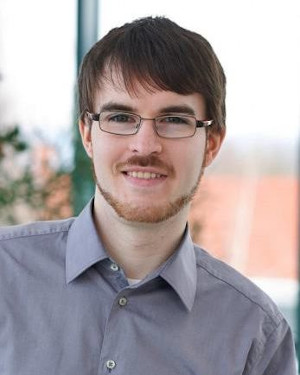}
	\end{wrapfigure}
	\textbf{Dipl.-Ing. Sebastian Gottschall} studied electrical engineering at the Technische Universit{\"a}t Dresden. He is working the team industrial cleaning technologies at the Fraunhofer IVV since 2019. Here he puts the focus on digitalization solutions including data collection and processing but also on cleaning sensors.
\end{minipage}

\begin{minipage}[t][1.2in]{\textwidth}
	\begin{wrapfigure}{l}{1.1in}
		\includegraphics[width=0.88in,height=1.1in,clip,keepaspectratio]{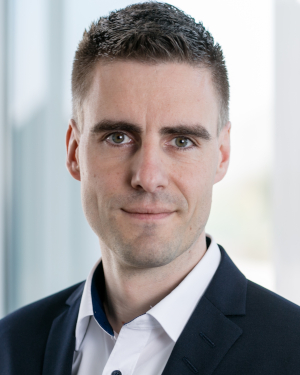}
	\end{wrapfigure}
	\textbf{Dr.-Ing. Lukas Oehm} graduated in mechanical engineering in 2010 and received his PhD in 2017 from Technische Universit{\"a}t Dresden entitled  »Joining of Polymeric Packaging Materials with High Intensity Focused Ultrasound«. He has been working as research assistant at Fraunhofer Institute for Process Engineering and Packaging IVV since May 2017. Since October 2018, he is group leader for Digitization and Assistance Systems. His research interests are in the field of product safety issues and process efficiency in food production as well as assistance systems for machine operators.
\end{minipage}

\endgroup

\end{document}